\documentclass[12pt,english]{article}
\usepackage[T1]{fontenc}
\usepackage[latin1]{inputenc}
\usepackage{amsmath}
\usepackage{graphicx}
\usepackage{cite}
\makeatletter

\makeatletter

\usepackage{amsfonts}

\usepackage{latexsym}

\usepackage{epsfig}

\makeatother

\usepackage{babel}
\makeatother

\begin{document}
\newcommand{\de}{\delta} \newcommand{\ga}{\gamma} \newcommand{\e}{\epsilon}
\newcommand{\ot}{\otimes} \newcommand{\be}{\begin{equation}
\end{equation}
 } \newcommand{\ee}{{equation}} \newcommand{\ba}{\begin{array}{c}
 \end{array}} \newcommand{\ea}{{array}} \newcommand{\beq}{\begin{equation}
\end{equation}
 } \newcommand{\eeq}{{equation}} \newcommand{\tmod}{{\cal T}} \newcommand{\amod}{{\cal A}}
\newcommand{\bemod}{{\cal B}} \newcommand{\cmod}{{\cal C}} \newcommand{\dmod}{{\cal D}}
\newcommand{\hmod}{{\cal H}} \newcommand{\s}{{\scriptstyle }} \newcommand{\tr}{{\rm tr}}
\newcommand{\einsop}{{\bf 1}} \def\R{\overline{R}} \def\doa{\downarrow}
\def\dag{\dagger} \def\ve{\epsilon} \def\si{\sigma} \def\ga{\gamma}
\def\nn{} \def\le{\langle} \def\re{\rangle} \def\lt{} \def\rt{}
\def\dwn{\downarrow} \def\up{\uparrow} \def\dag{\dagger} \def\bea{\begin{eqnarray}
\end{eqnarray}
 } \def\eea{{eqnarray}} \def\p{\tilde{p}} \def\q{\tilde{q}} \def\H{\overline{H}}
\newcommand{\reff}[1]{eq.~(\ref{#1})}

\title{Classical and quantum analysis of a hetero-triatomic molecular Bose-Einstein condensate model}

\author{A.P.Tonel$^{1}$,C.C.N. Kuhn$^2$,G. Santos$^3$,A. Foerster$^2$,I. Roditi$^4$ and Z.V.T. Santos$^4$  \vspace{1cm}
 \\
 $^{1}$CCET da Unipampa, Bag\'e, RS - Brazil\\
 $^{2}$Instituto de F\'{\i}sica - UFRGS, Porto Alegre, RS - Brazil\\
 $^{3}$Departamento de F\'{\i}sica - UFS, S\~ao Crist\'ov\~ao, SE - Brazil \\
 $^{4}$Centro Brasileiro de Pesquisas F\'{\i}sicas - CBPF, Rio de Janeiro - Brazil}

\maketitle
\begin{abstract}
We investigate an integrable Hamiltonian modelling a hetero-triatomic-molecular Bose-Einstein condensate.
This model describes a mixture of two species of atoms in different proportions, which can combine to form a triatomic 
molecule. Beginning with a classical analysis, we determine the fixed points of the system. 
Bifurcations of these points separate the parameter space into different regions. Three distinct scenarios  
are found, varying with the atomic population imbalance. This result suggests the ground 
state properties of the quantum model exhibits a sensitivity on the atomic population imbalance, which is 
confirmed by a quantum analysis using different approaches, such as the ground-state expectation values,
the behaviour of the quantum dynamics, the energy gap and the ground state fidelity. 
\end{abstract}


\vfil\eject 

\section{Introduction}
The experimental achievement that led to  the Bose-Einstein condensates (BECs), using
dilute alkali gases at ultracold temperatures \cite{early,angly}, induced a substantial effort 
dedicated to the understanding of new properties of BECs. In particular, the development of the 
techniques used in the production and manipulation of ultracold atoms and molecules \cite{Heinzen} 
has opened the way to a new field, the "chemistry" of ultracold systems, i.e   where the atomic constituents 
of the dilute gas may recombine forming molecules.  Such molecular BEC compounds  have been obtained by different 
techniques \cite{mol}, for instance,  by Feshbach resonances \cite{feshbach,catani5,catani55} 
or photoassociation \cite{photo}. 
There can also occur atom-molecule interactions that must be at least three-body in nature \cite{herbig}, bringing 
up new stimulus and challenges to our physical understanding.  Experimental  evidences for three-body 
recombinations \cite{weberPRL}  as well as for Efimov states \cite{EfimovEvidence} provide a physical 
ground and stimulus for the search of triatomic molecular BECs and for the investigation of their 
theoretical aspects, which is our main interest. 

>From the theoretical point of view, ultracold atomic and molecular systems are characterized by 
their large quantum fluctuations. In this sense, it becomes relevant the search for exactly solvable 
models  describing atomic and molecular BEC. Indeed this has become a very active field of 
research \cite{jon1,jonjpa,jonguan,key-3,dukelskyy,Ortiz,Kundu,eric5,jing}, and the experimental 
relevance of these models is currently a very active research subject \cite{drummond1}. 
Those solvable models are expected to provide a significant impact in this area, a view that 
has been promoted in \cite{hertier, batchelor}.
Increasing evidence and recent results show that multi-atomic systems may be interesting  
and relevant for ultracold atomic-molecular in Bose-Einstein condensates. 
A significant question in this context is whether more complex ultracold molecules 
could be created than simple dimers \cite{grimmnew}. Also, due to the more 
sophisticated nature of the control of the interatomic interactions, in the case of triatomic molecules, 
one expects a rich quantum phase structure.  Indeed, very recent experimental results confirm the 
existence of heteroatomic bosonic trimers in ultracold mixtures \cite{cataniprivate} which provide 
us with additional motivation to pursue the present investigation.

In this paper we analyze an integrable model describing a hetero-triatomic molecular Bose-Einstein
 condensate where atomic BECs can combine (in different proportions)
to produce a compound
 with two atoms of the same kind and a third one of a different
species. Our model, that has been shown to be solvable in \cite{jpa}, includes besides the interconversion of atoms to molecules and 
vice-versa, a linear interaction corresponding as the external potential and a bilinear interaction corresponding 
the scattering between atoms-atoms, atoms-molecules and molecules-molecules.
We start our analysis of this model by a classical treatment where we obtain its phase space determining in particular 
the fixed points. We find that for certain coupling parameters bifurcation of the fixed points occurs, and we can 
determine a parameter space diagram  which classifies the found fixed points. This diagram is determined for the 
imbalance of the number of atoms which allows us to classify it in three distinct cases. Specifically, when the 
imbalance is equal zero or negative there is a spontaneous appearance of additional boundaries in the parameter 
space (three for the zero case and two for the negative case), some of which can be identified with bifurcations 
of the minimum of the classical Hamiltonian. We also perform a quantum analysis, where we study the quantum dynamics 
and compare with the classical results. Here we are interested in studying the ground state of the model, because 
as actual systems are in ultracold temperature some insight can be obtained from the ground state. Furthermore as 
pointed out before the presence of large quantum fluctuations make it interesting to look for the phase structure 
at zero temperature, the quantum phase transitions. In our case we are able to look for signatures of quantum phase 
transition. Here we use two definitions, energy gap and ground state fidelity in order to find a quantum phase 
pre-transition, a term that will be explained later. We observe that the critical points are pinning down in 
completely agreement with the classical analysis.

The paper is organized as follows. In section 2 we present our integrable  model. Section 3 is devoted to the 
classical analysis where the parameter diagram are obtained. Section 4 is devoted to the quantum analysis where 
we show the quantum dynamics and a study about quantum phase pre-transition. Section 5 is devoted to our 
conclusions.

\section{The model}

Let us consider the following Hamiltonian describing a hetero-triatomic-molecular Bose-Einstein
condensate with two identical species of atoms, denoted by $a$ which can  be combined to a different type of atom, 
denoted by $b$, to produce a molecule labelled by $c$. In terms of canonical
creation and annihilation operators $\{a,b,c,a^{\dagger},b^{\dagger},c^{\dagger}\}$ satisfying the usual 
commutation relations $[a,a^{\dagger}]=I$, etc., the Hamiltonian reads

\begin{eqnarray}
H =U_{aa}N_{a}^{2}+U_{bb}N_{b}^{2}+U_{cc}N_{c}^{2}+U_{ab}N_{a}N_{b}+U_{ac}N_{a}N_{c}+U_{bc}N_{b}N_{c}+\\  \nonumber
+\mu_{a}N_{a}+\mu_{b}N_{b}+\mu_{c}N_{c}+\Omega(a^{\dag}a^{\dag}b^{\dag}c+c^{\dag}baa).
\label{ham}
\end{eqnarray}

\noindent The parameters $U_{ij}$ describe $S$-wave scattering, $\mu_{i}$
are external potentials and $\Omega$ is the amplitude for interconversion
of atoms and molecules. $N_{i}$ are the number operators, i.e
$N_{a}=a^{\dagger}a$ is the number of atoms type $a$, $N_{b}=b^{\dagger}b$ is the number of atoms type $b$ and $N_{c}=c^{\dagger}c$ is the number of molecules.

The Hamiltonian acts on the Fock space spanned by the (unnormalized) vectors
\begin{equation}
|N_{a};N_{b};N_{c}\rangle=(a^\dagger)^{N_a}(b^\dagger)^{N_b}(c^\dagger)^{N_c}|0\rangle,  
\label{fock}
\end{equation}

\noindent where $|0\rangle$ is Fock vacuum.

The Hamiltonian above has two independent conserved quantities
$$N=N_a+N_b+3N_c, \;\;\;\;\;\;\;\;\; J=N_a-2N_b,$$ where $N$ is the total number of atoms and $J$ is the atomic imbalance. It is convenient to introduce $k=J/N$, as the fractional atomic imbalance. Since there are three degrees of freedom and three conserved quantities, the model is integrable. More details about the integrability of this model, using the Bethe ansatz method, can be found in \cite{jpa}.
In what follows we will investigate this model in detail. Below we begin with a classical analysis of the model and determine the fixed points of the system.

\section{Classical analysis}

Let $N_{j},\,\theta_{j},\, j=a,\, b,\, c,$ be quantum variables satisfying
the canonical relations \[
[\theta_{j},\,\theta_{k}]=[N_{j},\, N_{k}]=0,~~~~~[N_{j},\,\theta_{k}]=i\delta_{jk}I.\]
 We make a change of variables from the operators $j,\, j^{\dagger},\, j=a,\, b,\, c,$
to a number-phase representation through \[
j=\exp(i\theta_{j})\sqrt{N_{j}},\;\;\;\;\;\;\; j=a,\, b,\, c,\]
 such that the canonical commutation relations are preserved.
We perform an additional change of variables \[
z=\frac{1}{N}(N_{a}+N_{b}-3N_{c}),\]
 \[
\theta=\frac{N}{6}(2\theta_{a}+\theta_{b}-\theta_{c}),\]
 such that $z$ and $\theta$ are canonically conjugate variables;
i.e. \[
[z,\,\theta]=iI.\]
 In the limit of large $N$ we can approximate the (rescaled)
Hamiltonian by

\begin{equation}
H=\frac{4\Omega N^{2}}{36}[\lambda z^{2}+2(\alpha-\lambda)z+\beta+(z+c_{+})\sqrt{(z+c_{-})(1-z)}\cos(\frac{6\theta}{N})],
\label{hamc}
\end{equation}

\noindent where we have defined
\begin{eqnarray*}
\lambda & = & \Delta(4U_{a}+U_{b}+U_{c}+2U_{ab}-2U_{ac}-U_{bc}),\\
\alpha & = & \Delta[4(c_{+}+1)U_{a}+(c_{-}+1)U_{b}+(c_{+}+c_{-}+2)U_{ab}-(1+c_{+})U_{ac}-(1+c_{-})\frac{U_{bc}}{2}\\
 & + & \frac{3}{N}(2\mu_{a}+\mu_{b}-\mu_{c})],\\
\beta & = & \Delta[4U_{a}c_{+}^{2}+U_{b}c_{-}^{2}+U_{c}+2U_{ab}c_{+}c_{-}+2U_{ac}c_{+}+U_{bc}c_{-}\\
 & + & \frac{6}{N}(2\mu_{a}c_{+}+\mu_{b}c_{-}+\mu_{c})],
\end{eqnarray*}

\noindent with 
\[
c_{-}=1-2k,\;\;\;\; c_{+}=1+k,\;\;\;\;\Delta=\frac{1}{4\Omega},\ \]
\[
k=\frac{J}{N},\;\;\;\; k\in[-2,1].\]

We now regard (\ref{hamc}) as a classical Hamiltonian and
investigate the fixed points of the system. The first step is to find the
Hamilton's equations of motion which yields 

\begin{eqnarray}\label{eh}
\frac{dz}{dt} & = & \frac{\partial H}{\partial\theta}=-\frac{4\Omega N}{6}(z+c_{+})\sqrt{(z+c_{-})(1-z)}\sin\left(\frac{6\theta}{N}\right),\nonumber \\
\frac{d\theta}{dt} & = & -\frac{\partial H}{\partial z}=\frac{4\Omega N^{2}}{36}[2\lambda z+2(\alpha-\lambda)\\
 & + & \frac{2(z+c_{-})(1-z)+(z+c_{+})(1-z)-(z+c_{+})(z+c_{-})}{2\sqrt{(z+c_{-})(1-z)}}\cos\left(\frac{6\theta}{N}\right)].\nonumber 
\end{eqnarray}

 The fixed points of the system are determined by the condition 
\begin{equation}
\frac{\partial H}{\partial\theta}=\frac{\partial H}{\partial z}=0.\label{fixed}
\end{equation}
Due to periodicity of the solutions, below we restrict to $\theta\in[0,\, N\pi/3]$. It is convenient to define the functions:
\[
f(z)=\lambda z+\alpha-\lambda,\]

\[
g(z)=-\frac{2(1-z)(z+c_{-})+(1-z)(z+c_{+})-(z+c_{+})(z+c_{-})}{4\sqrt{(1-z)(z+c_{-})}}.
\]

\noindent Note that the domain of $g(z)$ is $z\in [-1,1)$ if $k\in [-2,0]$ and $z\in (2k-1,1)$ if $k\in (0,1)$.

\begin{figure}[ht]
\begin{centering}
\includegraphics[scale=0.35]{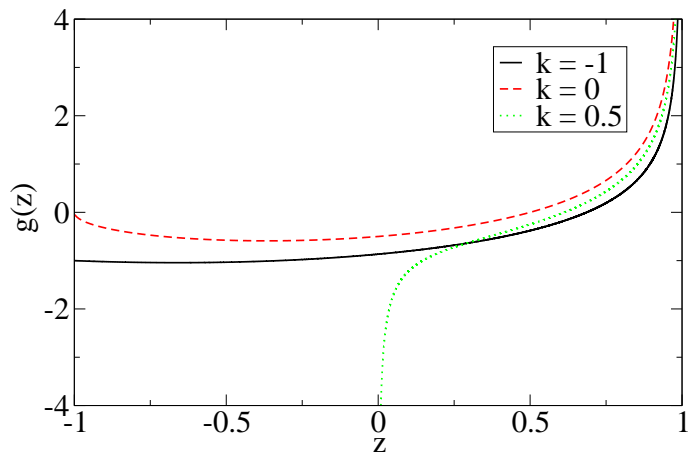}
\vspace{1.0cm}
\par\end{centering}
\caption{ (Color online) The behaviour of the function $g(z)$ for three different values of $k$.}
\label{fig:1}
\end{figure}

We observe that the fractional atomic imbalance $k$ plays an important role in the behaviour of the $g(z)$  function. For $k \leq 0$, $g(z)$ is divergent only at $z=1$, while for the case of $k > 0$, $g(z)$ is divergent at $z = 2k-1$ and $z = 1$. Since $k$ affects the domain and the shape of the function $g(z)$, this property will affect the type of solutions of (\ref{fixed}). In Fig. \ref{fig:1} we illustrate the behaviour of the function $g(z)$ for different values of $k$. It is, in fact, necessary to treat the cases of $k < 0$, $k=0$ and $k > 0$ separately.

\vspace{1.0cm}

\subsection{Negative case : $-2 \leq k < 0 $}
Here the domain of $g(z)$ is $z \in [-1,1)$ and $g(z)$ is divergent at $z=1$, but finite at $z=-1$. This leads to the following classification for the solutions of (\ref{fixed}):

\begin{itemize}
\item $\theta=0$ and $z$ is a solution of

\begin{equation}
f(z)= g(z),
\label{sol1}
\end{equation}
\noindent which can admit zero, one or two solutions.

\item  $\theta=N\pi/6$ and $z$ is a solution of

\begin{equation}
f(z)= - g(z),
\label{asol1}
\end{equation}
\noindent which can admit zero, one or two solutions.

\item $z=-c_+$, which vanishes the first equation of (\ref{eh}) and reduces the second eq. of (\ref{eh}) to the expression

\begin{equation}
\lambda = \frac{\alpha}{k+2} + \frac{\sqrt{-3k(k+2)}}{2(k+2)}\cos(\frac{6\theta}{N}),
\label{boundary1}
\end{equation}
\noindent such that $\theta$ is a solution of

\begin{equation}
\cos(\frac{6\theta}{N})=-\frac{2\sqrt{-3k(k+2)}}{3k}\left( \lambda - \frac{\alpha}{k+2}\right),
\label{valtheta}
\end{equation}
\noindent for which there are two solutions for $|\frac{2\sqrt{-3k(k+2)}}{3k}\left( \lambda - \frac{\alpha}{k+2}\right)| < 1$.

\end{itemize}

\subsection{Zero case: $ k = 0$}
Now we consider the case $k=0$, where the domain of $g(z)$ is $z \in (-1,1)$ and $g(z)$ is divergent at $z=1$, but finite at $z=-1$, similar to the previous case. This leads to the following classification for the general problem:

\begin{itemize}

\item $\theta=0$ and $z$ is a solution of

\begin{equation}
f(z)=g(z),
\label{sol3}
\end{equation}
\noindent which can admit zero, one or two solutions.

\item  $\theta=N\pi/6$ and $z$ is a solution of

\begin{equation}
f(z)= -g(z),
\label{asol3}
\end{equation}
\noindent which can admit zero, one or two solutions.

\item $z=-1$, which vanishes the first eq. of (\ref{eh}) and reduces the second eq. of (\ref{eh}) to the following linear equation between the coupling parameters  

\begin{equation}
\lambda =\frac{\alpha}{2},
\label{boundary}
\end{equation}
\noindent which can admit just one solution. This result is compatible with that obtained in the previous case by taking the limit $k  \rightarrow 0 $ in eq. (\ref{boundary1})

\end{itemize}

\subsection{Positive case: $0 < k\leq 1$}
In this case the domain of $g(z)$ is $z \in (2k-1,1)$ and $g(z)$ is divergent at both extremes of the interval, $z=2k-1$ and $z=1$. Now, a different scenario emerges, compared to the previous two cases. This leads to the following classification for the general problem:

\begin{itemize}
\item $\theta=0$ and $z$ is a solution of

\begin{equation}
f(z)=g(z),
\label{sol11}
\end{equation}
\noindent which can admit one, two or three solutions.

\item $\theta=N\pi/6$ and $z$ is a solution of

\begin{equation}
f(z)= - g(z),
\label{asol11}
\end{equation}
\noindent which can admit one, two or three solutions.
\end{itemize}

\begin{figure}[ht]
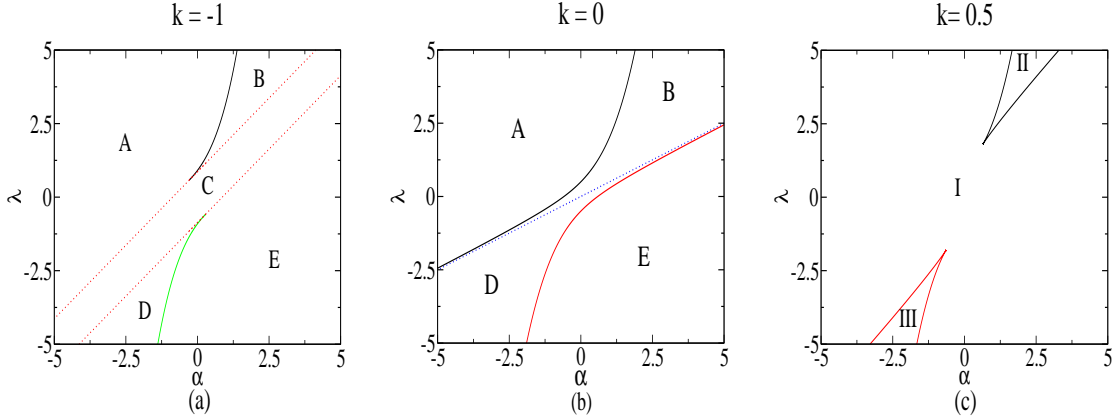

\begin{centering}
\epsfig{file=dpm1.eps,width=4.5cm,height=5.5cm,angle=0}\hspace{0.6cm}\epsfig{file=dpk0.eps,width=4.5cm,height=5.5cm,angle=0}\hspace{0.6cm}\epsfig{file=dpk05.eps,width=4.5cm,height=5.5cm,angle=0}
\par\end{centering}
\caption{(Color online) Parameter space diagram identifying the different types of solution for eq.(\ref{fixed}) for different values of $k=-1;0;0.5$. We observe (a) five distinct regions for the negative case; (b) four distinct regions for $ k = 0$; (c) three distinct regions for the positive case. In (a) the boundaries are given by $\lambda = (\alpha \mp g(-k-1))/(k+2)$ while in (b) it is given by $\lambda=\alpha/2$.}
\label{fig:2}
\end{figure}

We can collect all different types of solutions of eq. (\ref{fixed}) in a parameter diagram, dividing the parameter space into different regions, for each case of $k$ discussed above. For example, for the case of $k$ positive, to construct this diagram, we observe that the boundaries between each regions occur when $f$ is the tangent line to $\pm g$; i. e. for values of $\lambda$ and $\alpha$  such that

\begin{eqnarray*}
\lambda & = & \pm \frac{dg}{dz}|_{z_0},\\
f(z_0) &=& \pm g(z_0), \nonumber
\end{eqnarray*} 
\noindent for some $z_0$. This requirement determines the boundaries in the parameter space, which are depicted in Fig. \ref{fig:2}(c) for $k=0.5$.
\vspace{1.0cm}

As in the case of $k$ positive, we can determine the region boundaries in the parameter space for the other two cases. 
However, because of the existence of solutions of the form given by (\ref{boundary1}), which do not have an analogue 
for positive $k$, we see the appearance of new boundaries given by the conditions $\lambda = (\alpha \mp g(-k-1))/(k+2)$ 
for negative $k$ and $\lambda = \alpha/2 $ for $ k = 0 $. The boundaries in parameter space are illustrated in 
Fig. \ref{fig:2}(a) and Fig. \ref{fig:2}(b) for $ k=-1 $ and $ k=0 $, respectively. Notice that the two additional 
boundaries, which delimit region C, for  $k=-1 $ are reduced to a unique boundary for $ k=0 $, which is not present 
for $ k=0.5 $. Therefore, we have a different scenario for the parameter space diagram, depending if the fractional 
atomic imbalance $k$ is negative, zero or positive, as illustrated in Fig. \ref{fig:2}. 
Basically, we can summarize the typical behaviour of the parameter space diagram as follows: when $k$ 
is negative, the parameter diagram is divided in five regions: in region A there is no solution for $z$ 
when $\theta=0$ and one solution for z when $\theta=N\pi/6$. In region B there are two solutions for $z$ 
when $\theta=0$ and one solution for $z$ when $\theta=N\pi/6$. In region C there is one solution for $z$ 
when $\theta=0$, one solution for $z$ when $\theta=N\pi/6$ and two solutions for $\theta$ when $z=-k-1$. 
In region D there is one solution for $z$ when $\theta=0$ and two solutions for  $z$ when $\theta=N\pi/6$. 
In region E there is one solution for $z$ when $\theta=0$ and no solution for $z$ when $\theta=N\pi/6$. 
For the case $k = 0$,  region C disappears and the phase diagram is left with the four regions A, B, D, E 
discussed before. When $k$ is positive the diagram is divided in three regions: in region I there is one 
solution for $z$ when $\theta=0$ and one solution for $z$ when $\theta=N\pi/6$. In region II there are 
three solutions for $z$ when $\theta=0$ and one solution for $z$ when $\theta=N\pi/6$. In region III there is 
one solution for $z$ when $\theta=0$ and three solutions for $z$ when $\theta=N\pi/6$.
It is interesting to mention that the fractional atomic imbalance also plays an important role in hetero-diatomic 
molecular Bose-Einstein condensates \cite{eduardo, weiping,weiping5}.

\begin{figure}[ht]
\begin{centering}
\begin{tabular}{ccc}
 $(a)\;k= -1$ & $(b)\;k=0$ & $(c)\;k=0.5$ \\
\includegraphics[scale=0.45]{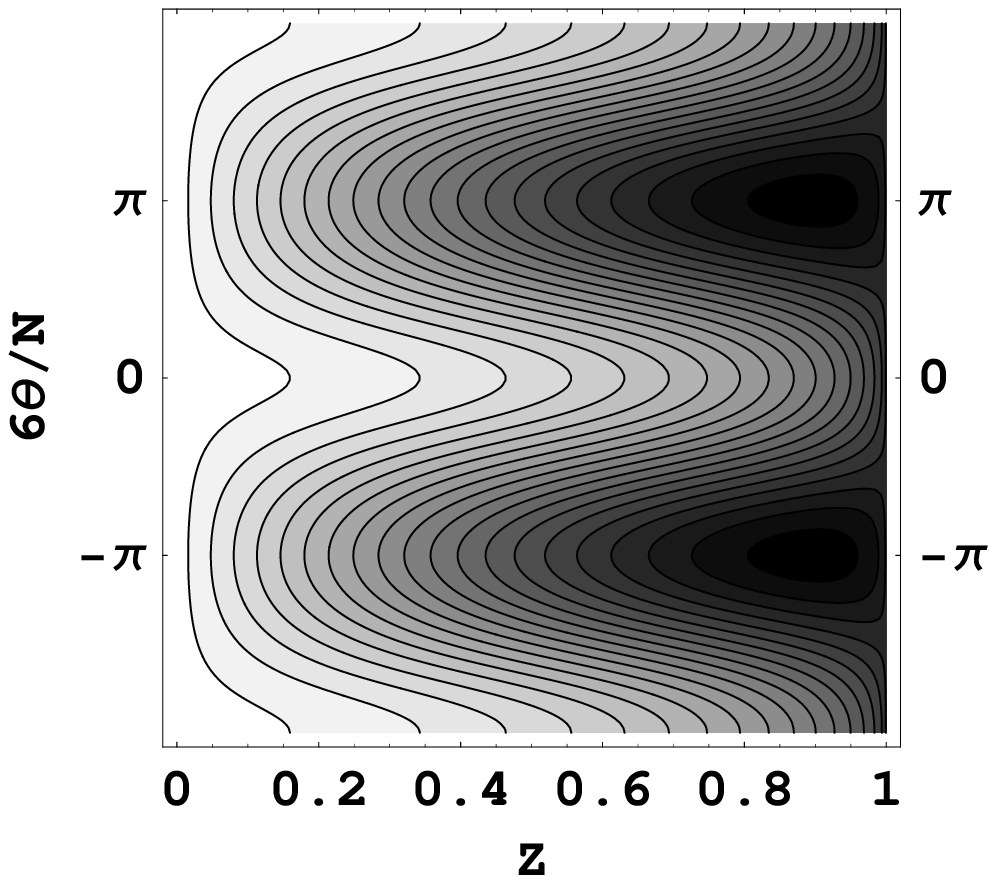}&\includegraphics[scale=0.45]{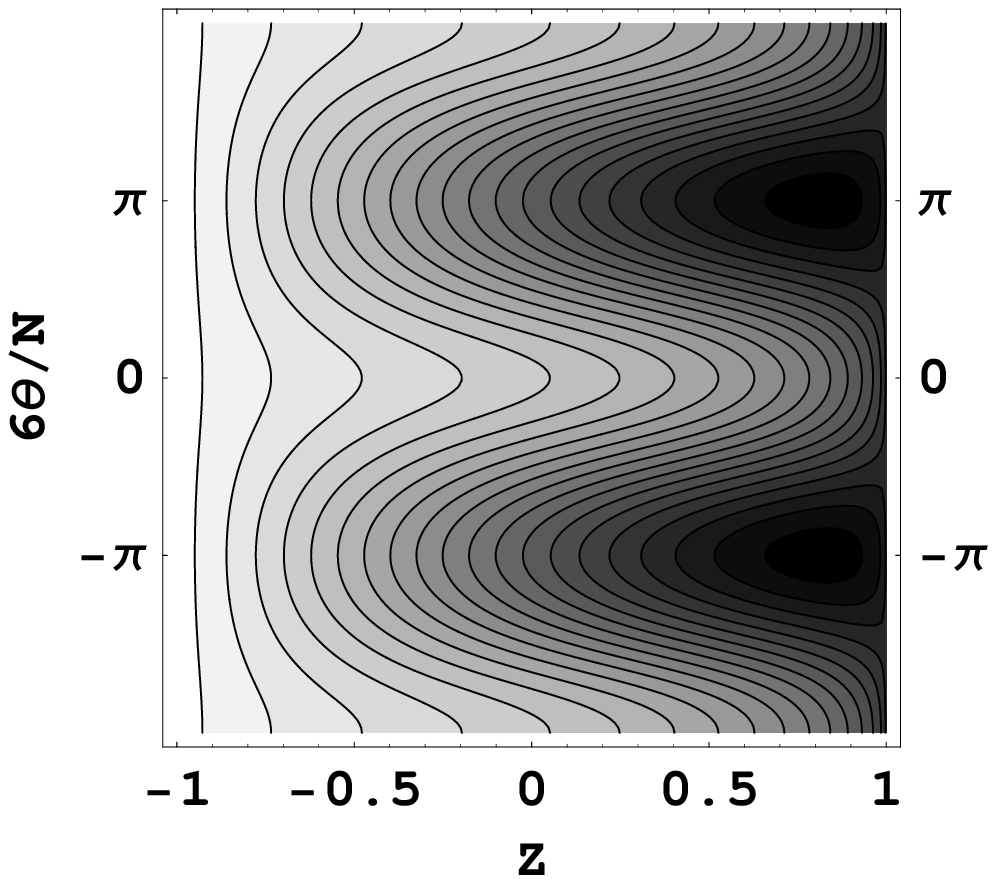}&\includegraphics[scale=0.45]{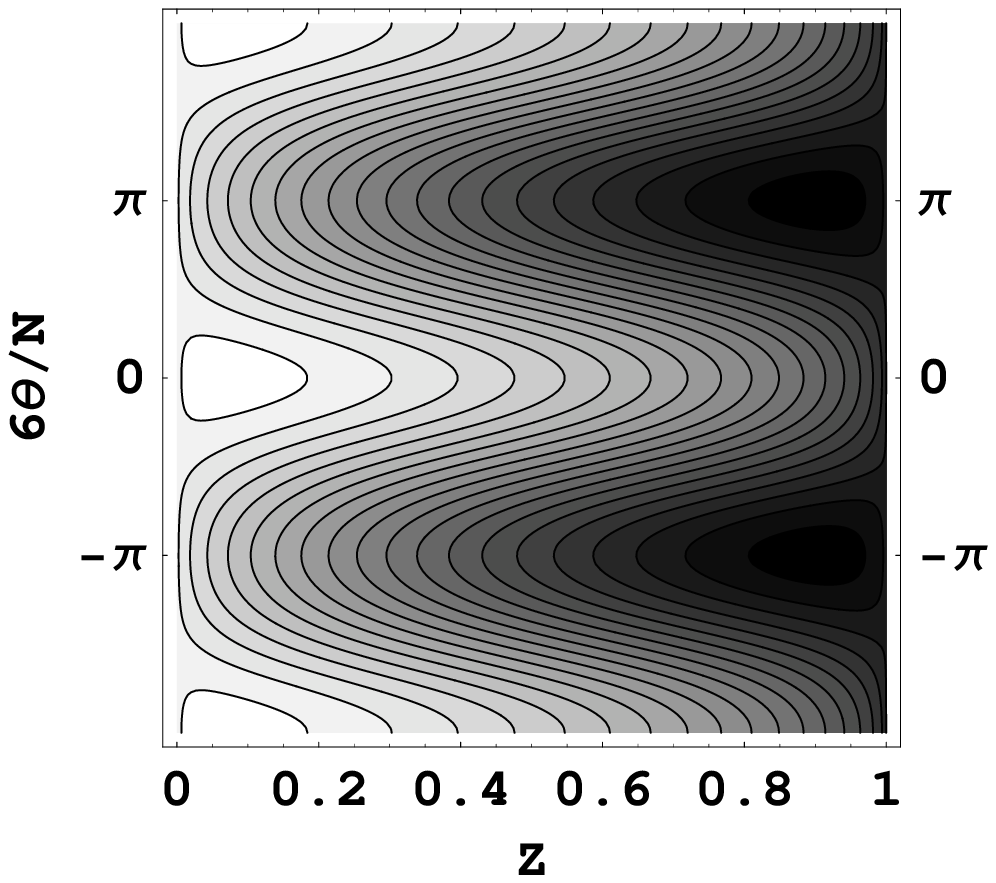} \\
\includegraphics[scale=0.45]{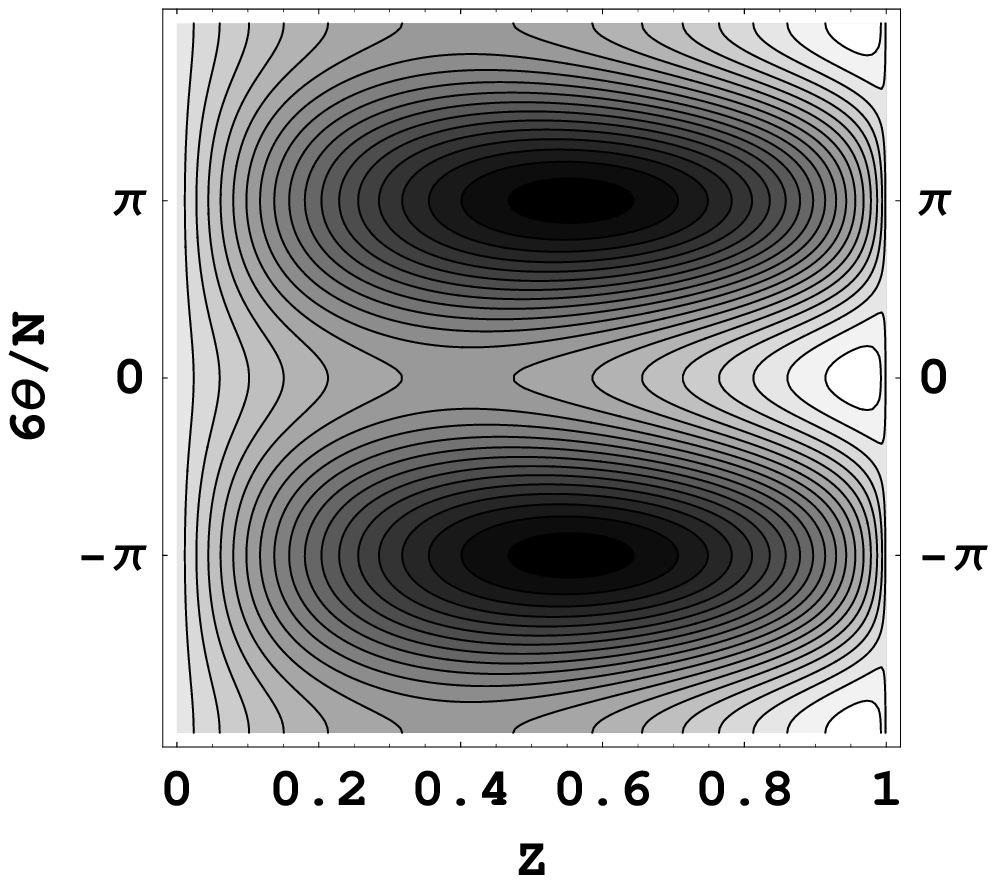}&\includegraphics[scale=0.45]{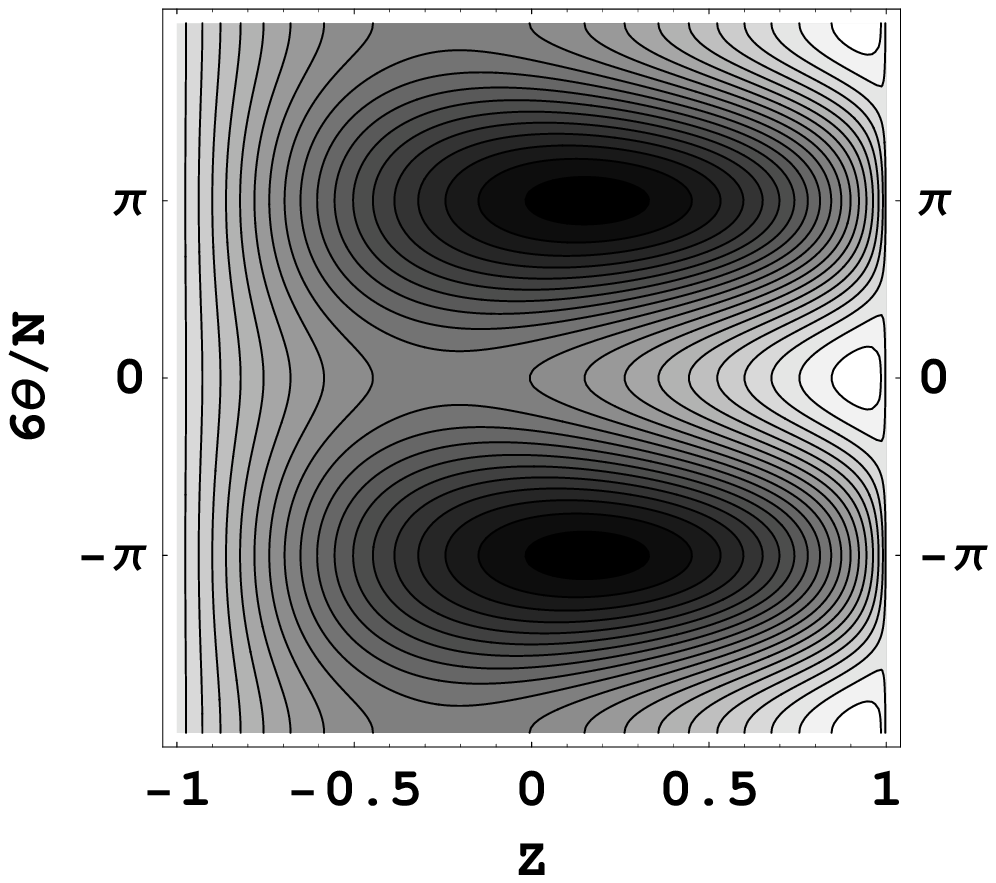}&\includegraphics[scale=0.45]{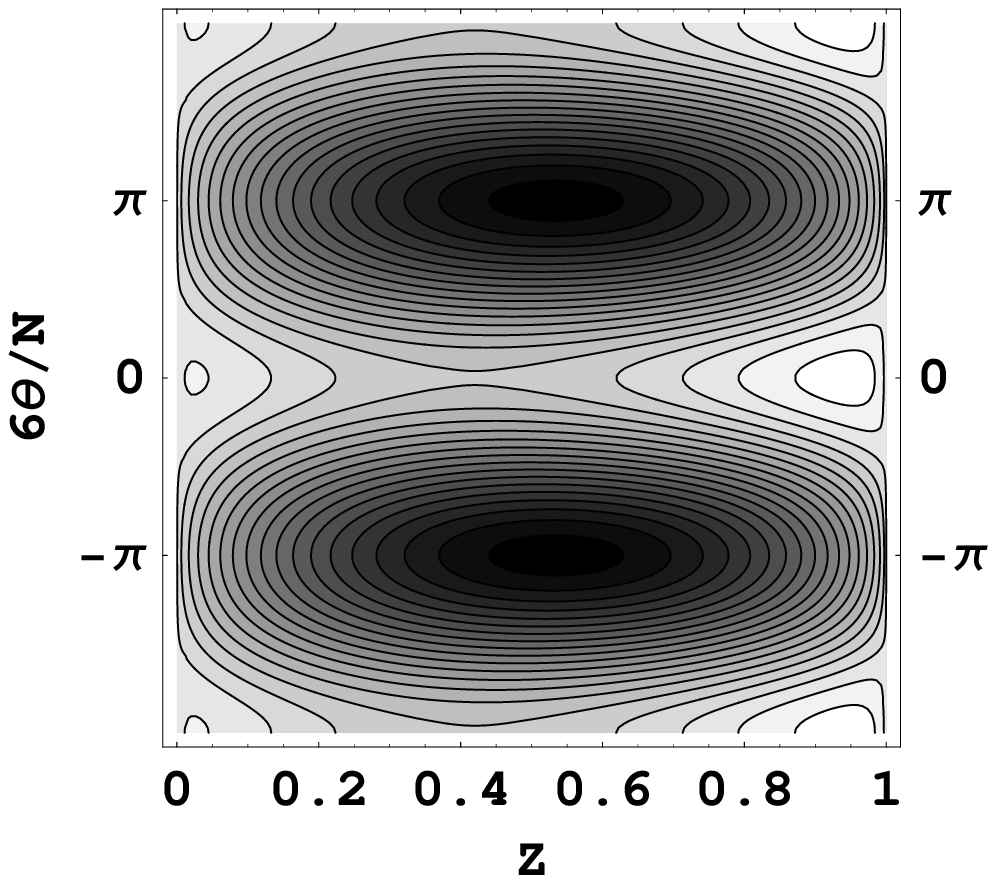} \\
\includegraphics[scale=0.45]{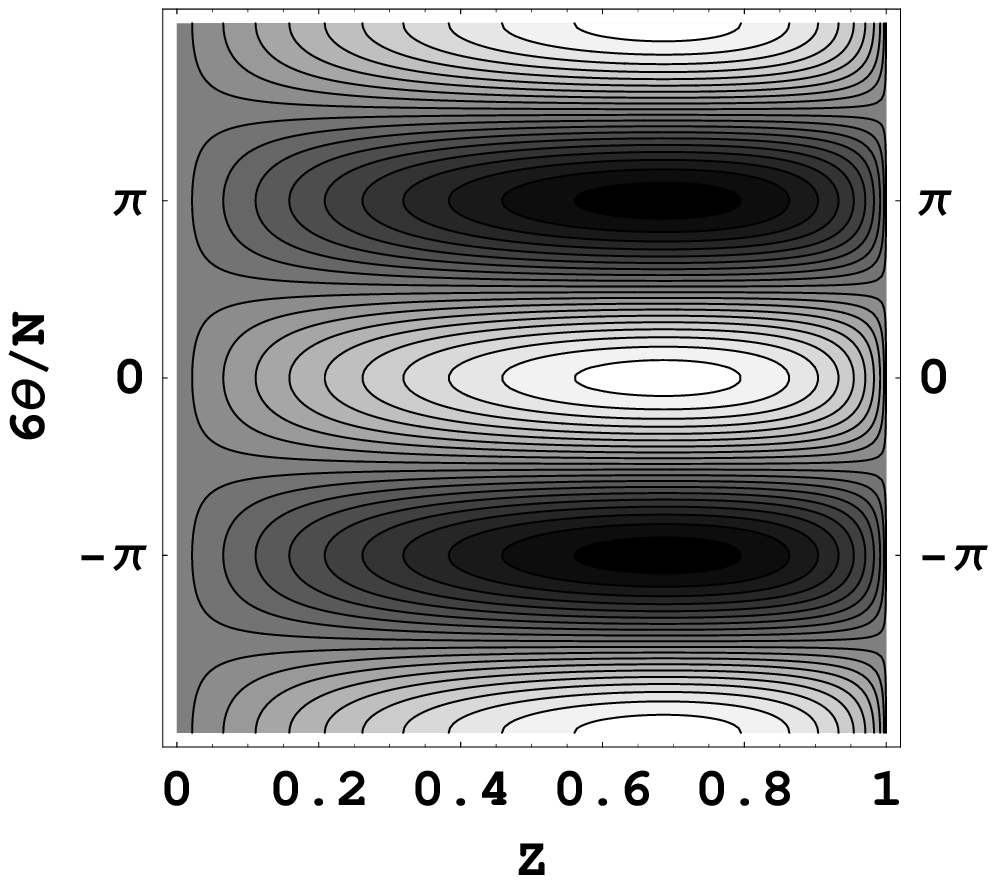}&\includegraphics[scale=0.45]{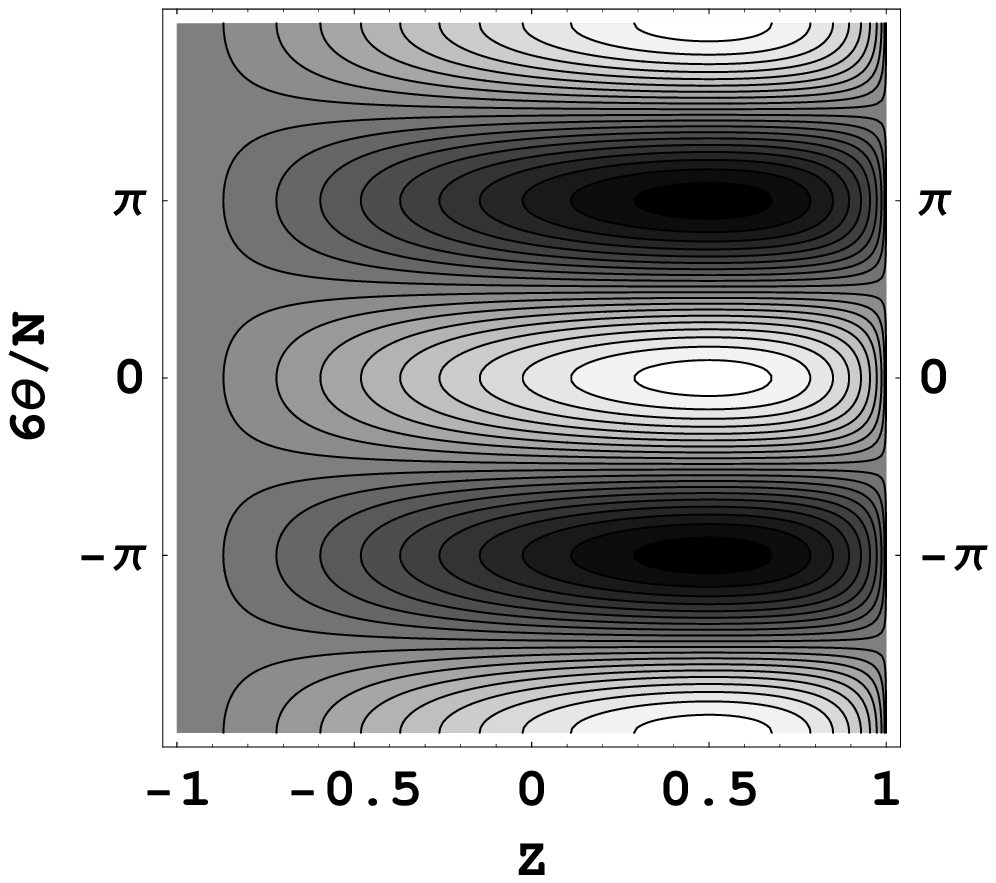}&\includegraphics[scale=0.45]{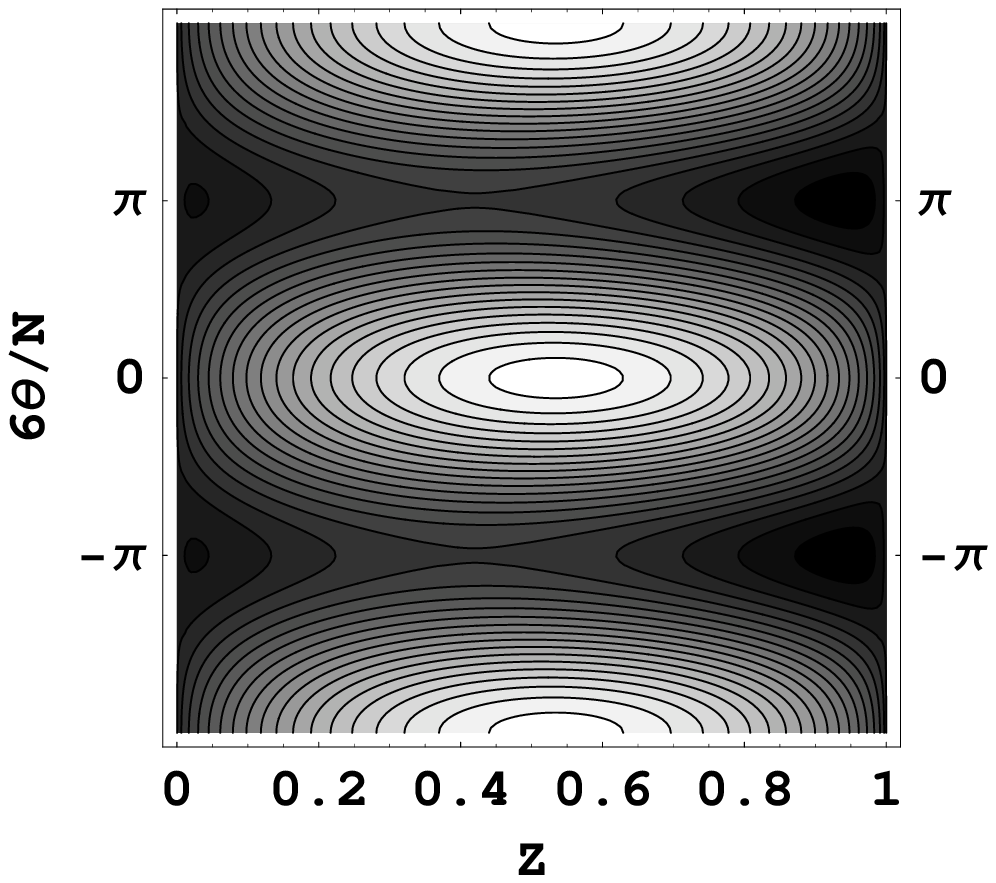} \\
\includegraphics[scale=0.45]{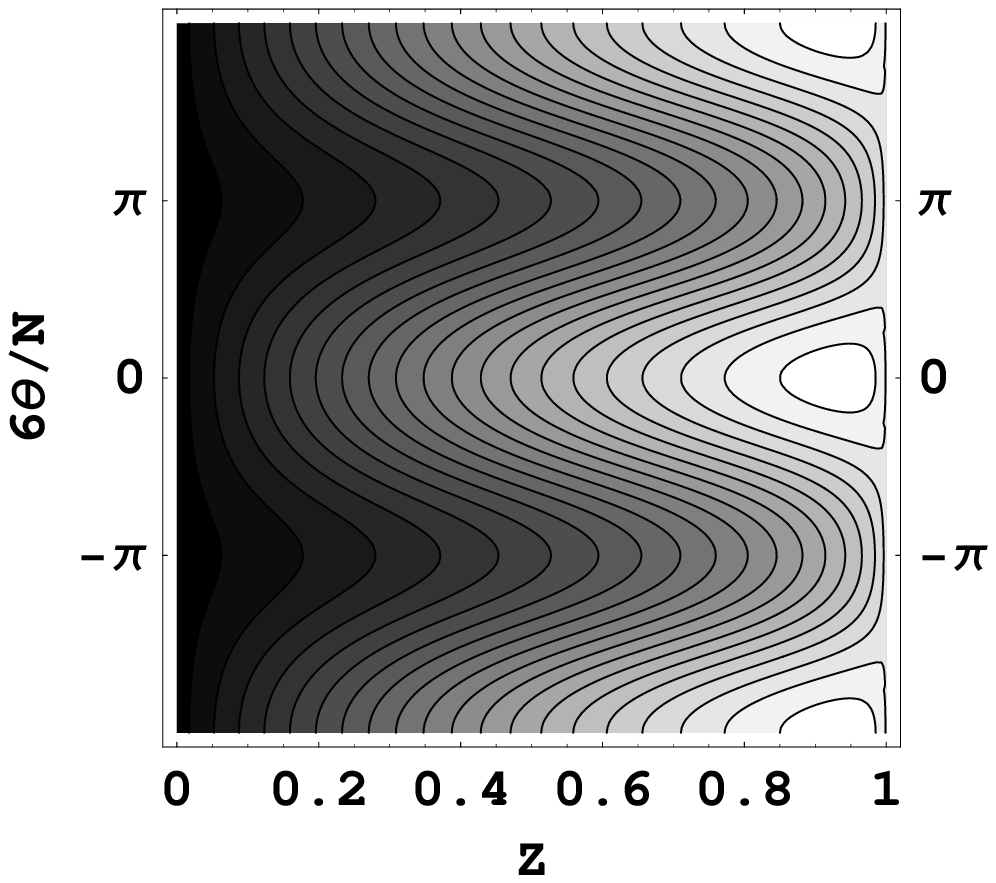}&\includegraphics[scale=0.45]{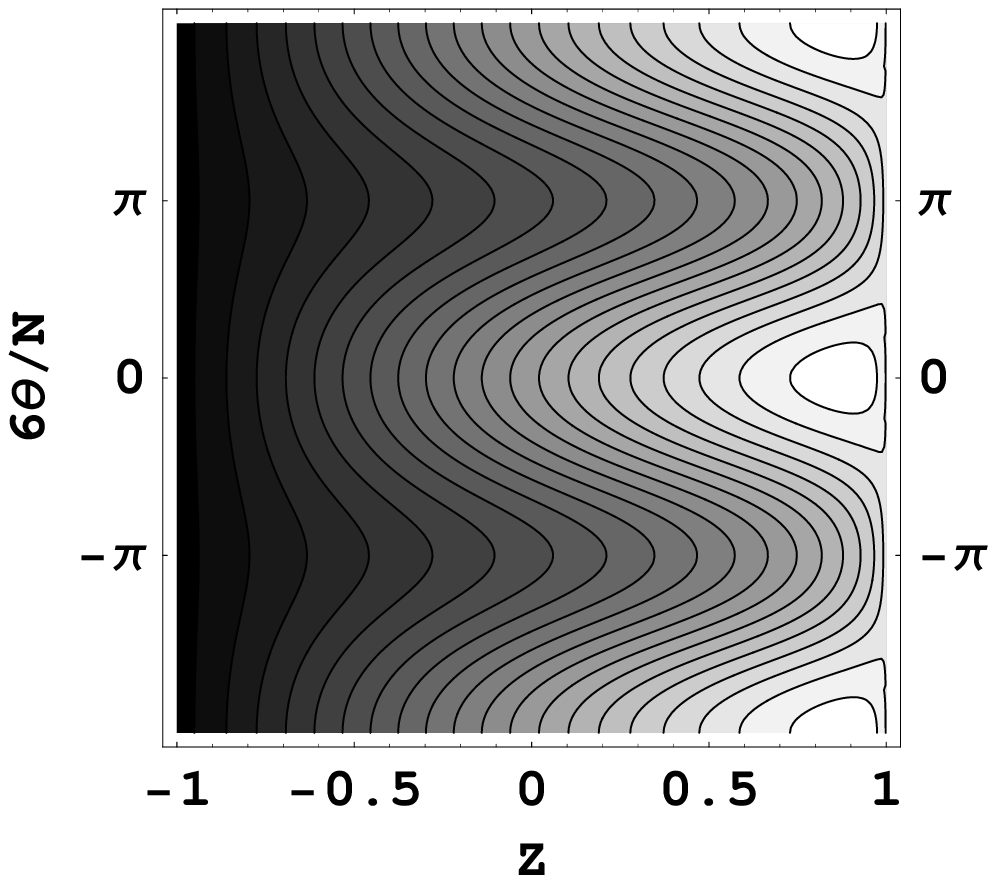}&\includegraphics[scale=0.45]{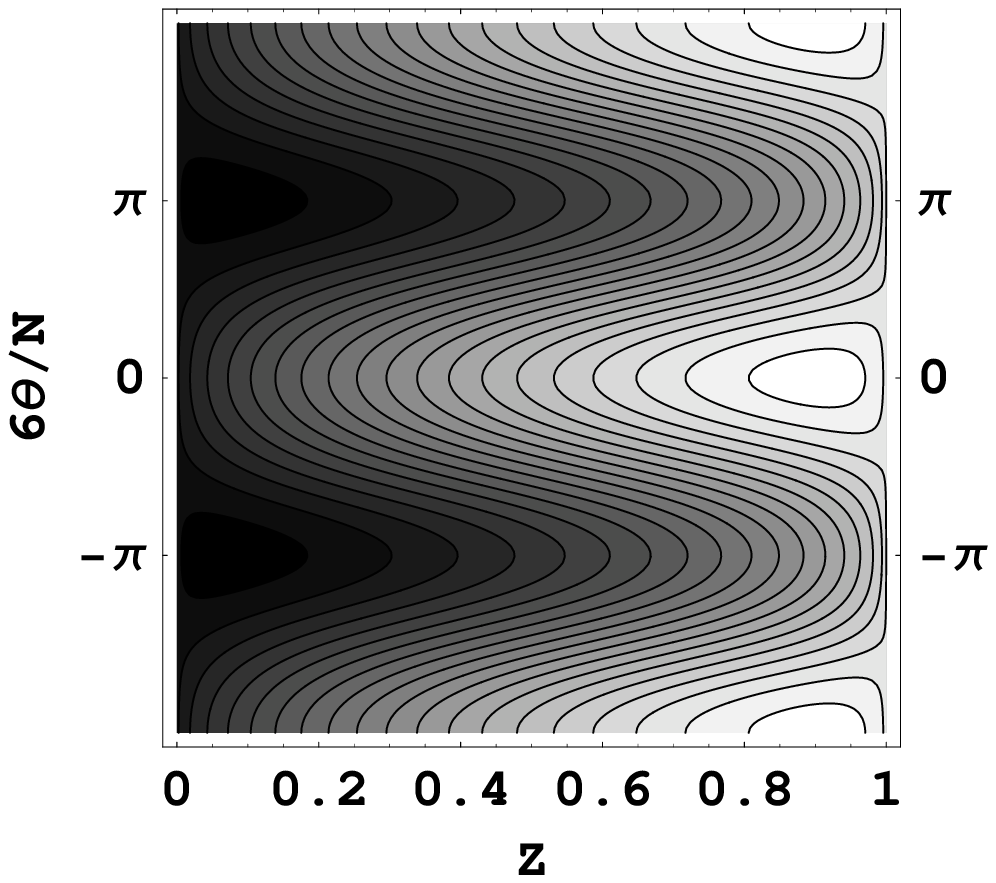} \\
\end{tabular}
\par\end{centering}
\caption{ Level curves for the Hamiltonian (\ref{hamc}), where the dark regions indicate lower values than the light regions. Here we
are using for: (a) $k=-1$ on the left $(\lambda,\alpha)=(0;-1.0),(5;2.5),(0,0)$ and $(0,1.5)$; (b) $k=0$ in  the middle $(\lambda,\alpha)=(0,-1),(2.5,2.5),(0;0)$ and $(0,1.5)$;(c) $k=0.5$ on the right $(\lambda,\alpha)=(0;-1.5),(5,2.5),(-5,-2.5)$ and $(0,1.5)$.}
\label{lc1} 
\end{figure}

To help  visualize the classical dynamics, it is useful to plot the level curves of the Hamiltonian (\ref{hamc}). 
Since the fixed points change the topology of the level curves, qualitative differences can be observed between the 
different regions.
The results are depicted in Fig. \ref{lc1} for $k=-1$ (on the left), $k=0$ (in the middle) and $k=0.5$ (on the right). 
For clarity, we use convenient intervals for $\theta$ and $z$.

\vspace{1.0 cm}

In Fig. \ref{lc1}(a) we show the level curves of the Hamiltonian (\ref{hamc}) for $k=-1$, illustrating the typical behaviour for regions A, B, C and E (from the top to the bottom). In region A there are local minima at $6\theta/N = \pm \pi$. Besides the minima at $6\theta/N = \pm \pi$, two additional fixed points (a maximum and a saddle point) are apparent in region B occurring at $\theta=0$. In region C there are minima at $6\theta/N = \pm \pi$ and for $\theta=0$ just one fixed point, a maximum. There are also saddle points when $z=0$. In region E just one fixed point, a maximum, occurs for $\theta=0$.

In Fig. \ref{lc1}(b) we show the level curves for $k=0$ for the same regions illustrated in the previous case, except that now instead of region C there is just one straight line separating regions B and D. The behaviour here is analogous to the previous case of negative $k$, with the emergence of a maximum (minimum) when passing from region A to B (E to C).

In Fig. \ref{lc1}(c) we present the level curves of the Hamiltonian (\ref{hamc}) for $k=0.5$, illustrating the behaviour of regions I, II, III and I (from the top to the bottom). In region I there is a maximal point at $\theta=0$ and a minima at $6\theta/N= \pm \pi$. Two additional fixed points, a saddle and a maximum occur, in region II at $\theta=0$, while two additional fixed points, a saddle and a minimum, occur in region III at $6\theta/N= \pm \pi$ compared to region I.

We observe that the pattern of the level curves is distinct for the cases of $k$ negative and zero compared to the positive case.

In the following sections we will conduct an analysis of the quantum Hamiltonian. We will focus our attention on the case $\lambda=0$, in this way the model has one effective parameter, $\alpha$. In particular we will establish that the bifurcation occurring at $(\alpha,\lambda)=(-g(-k-1),0)$ for the negative case and $(\alpha,\lambda)=(0.5,0)$ for $k=0$ can be seen to influence the ground state properties of the quantum system.

\section{Quantum analysis}
We now turn our attention to a quantum treatment of the model, to investigate the nature of the additional threshold couplings for the cases where the fractional atomic imbalance $k$ is negative and zero. In particular we analyze the Hamiltonian in the {\it no scattering limit} where $U_{ij}=0$ for all $i,j=a,b,c,$
 
\begin{equation}
H=\mu_{a}N_{a}+\mu_{b}N_{b}+\mu_{c}N_{c}+\Omega(a^{\dag}a^{\dag}b^{\dag}c+c^{\dag}baa).
\label{hamreduzido}
\end{equation}
 
 This simplifies substantially the Hamiltonian, however it remains sufficiently non trivial to enable us to gain an understanding of the quantum behaviour through the quantum dynamics, ground-state expectation value, gap and fidelity. The {\it no scattering limit} corresponds to the coupling $\lambda=0$ in the classical analysis of section $3$. With reference to Fig. \ref{fig:2} there are two threshold couplings when $k$ is negative and three threshold couplings for $k=0$. For the case of $k$ negative, one occurs at $(\alpha,\lambda)=(-g(-k-1),0)$, signifying the bifurcation of the global minimum of the Hamiltonian, while the other occurs at $(\alpha,\lambda)=(g(-k-1),0)$, signifying the bifurcation of the global maximum. For the specific example of $k=-1$, these thresholds are $(\alpha, \lambda)=(0.866,0)$ and $(\alpha, \lambda)=(-0.866,0)$, respectively. For the case $k=0$, there are three bifurcations at $(\alpha,\lambda)=(-0.5,0),(0,0),(0.5,0)$. The case $(\alpha,\lambda)=(-0.5,0)$ signifying the bifurcation of the global maximum, $(\alpha,\lambda)=(0.5,0)$ signifying the bifurcation of the global minimum while $(\alpha,\lambda)=(0,0)$ signifying the bifurcation of the saddle point. In contrast, there are no bifurcations along the line  $\lambda=0$ for the positive case. We focus our attention to the coupling $(\alpha,\lambda)=(0.866,0)$ for $k=-1$ and $(\alpha,\lambda)=(0.5,0)$ for $k=0$, as in these cases the bifurcation of the fixed point in phase space is associated with the ground state of the quantum system.

\subsection{Quantum dynamics}
In general the time evolution of any state is given by $|\Psi(t) \rangle = U(t)|\phi \rangle$, 
where $U(t)$ is the temporal operator $U(t)=\sum_{m=0}^{M}|m\rangle \langle m|\exp(-i E_{m} t)$,  
$|m\rangle$ is an eigenstate with energy $E_{m}$ and $|\phi \rangle$ represents  the initial state with $N=N_a+N_b+3N_c$. We adopt the method of directly diagonalizing the Hamiltonian (\ref{hamreduzido}) as done in \cite{our,ours} and compute the expectation value of $z(t)$ through.

\begin{equation}
\langle z(t)\rangle=\frac{1}{N}\langle \Psi (t)|N_a+N_b-3N_c|\Psi (t)\rangle.
\label{expectation}
\end{equation}

In our analysis, for fixed  total number of atoms $N$ and fixed atomic imbalance $J$, we will use the initial state configuration $|0,-J/2,(2N+J)/6\rangle$ for the cases where $k$ is negative and zero and  $|J,0,(N-J)/3\rangle$ for the case where $k$ is positive.
We therefore compare the three cases of the quantum dynamics, with fractional atomic imbalance negative, zero and positive.

\begin{figure}[ht]
\begin{center}
\epsfig{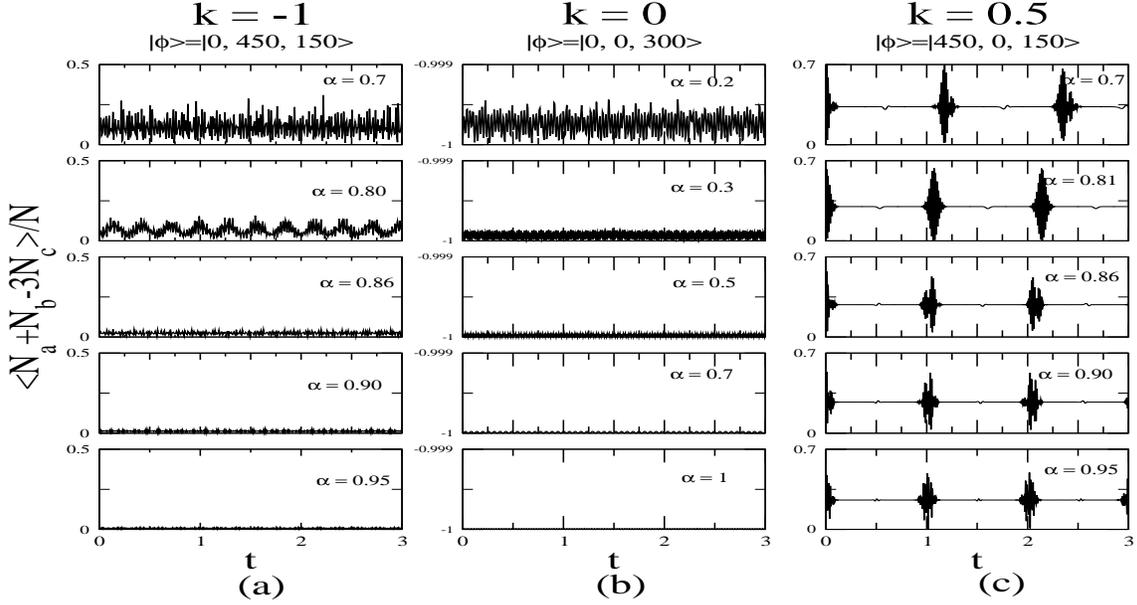}
\end{center}

\caption{Time evolution of the expectation value of $z$ for the Hamiltonian (\ref{hamreduzido}) with $N=900$, for (a) $k=-1$ and initial state $|0,450,150\rangle$. We are using natural units. The oscillations are largely irregular with significantly decreasing amplitude as the point at $\alpha=0.866$ is crossed. This point corresponds to the boundary at $(\alpha,\lambda)=(0.866,0)$ between regions C and E as shown in Fig. \ref{fig:2}(a); (b) $k=0$ and initial state $|0,0,300\rangle$. A similar behaviour occurs as the point at $\alpha=0.5$ is crossed. This point correspond to the boundary at $(\alpha,\lambda)=(0.5,0)$ as shown in Fig \ref{fig:2}(b); (c) $k=0.5$ with initial state $|450,0,150\rangle$. The oscillations display collapse and revival behaviour with smoothly decreasing amplitude. Here there is no abrupt behaviour, indicative of the fact there is no boundary at $\lambda=0$ in Fig. \ref{fig:2}(c).}
\label{dynamics}
\end{figure}

Results of the expectation value for $z$ are shown in Fig. \ref{dynamics} for the 
cases of $k=-1$, $0$ and $0.5$. We are using $N=900$ and $J=-900; 0; 450$ for $k=-1; 0; 0.5$, respectively. 
We fix the parameter $\Omega=1$ and use $\mu_c$ as the variable coupling parameter. 
In terms of the classical variables, this corresponds to vary the parameter $\alpha$. 
The qualitative differences are quite apparent. In the case of $k=-1$, Fig. \ref{dynamics}(a), we find 
that for $\alpha<0.866$ there are irregular oscillations in $z$. Similar behaviour occurs for $\alpha<0.5$ for $k=0$, 
Fig. \ref{dynamics}(b). As the coupling parameter $\alpha$ is increased across the threshold value at $\alpha=0.866$, 
for $k=-1$ and $\alpha=0.5$ for $k=0$, the transition to localized oscillations is significant in cases (a) and (b). 
By comparison the dynamics in Fig. \ref{dynamics}(c) for $k=0.5$ show a collapse and revival of oscillations.

\vspace{1.0 cm}

\subsection{Ground state expectation values}

\begin{figure}[ht]
\begin{center}
\epsfig{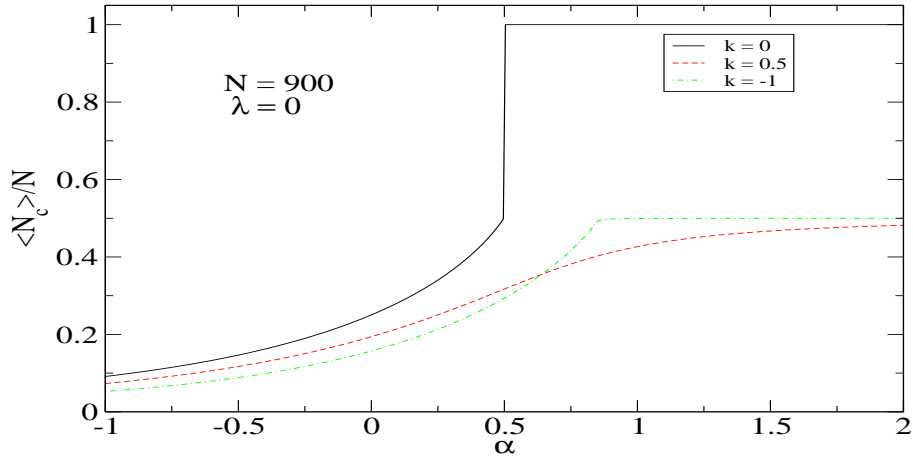}
\end{center}
\caption{(Color online) Normalized ground-state expectation value of the molecular number operator $\langle N_c\rangle$ versus the coupling parameter $ \alpha $ for the three different cases $k=-1$, $0$ and $0,5$. Here we are using $\Omega=1$ and $N=900$. For the cases $k=-1$ and $k=0$ there is an abrupt change in the expectation value $3\langle N_c\rangle/N$ as the threshold coupling $\alpha=0.866$ (for $k=-1$) and $\alpha=0.5$ (for $k=0$) is reached. In contrast, for $k=0.5$, the expectation value $3\langle N_c\rangle/N$ increases smoothly with $\alpha$, not exhibiting any abrupt behaviour.}
\label{figexpectation}
\end{figure}

Now using the equation (\ref{expectation}), we compute the normalized  ground-state expectation value $3\langle N_c \rangle/N$ for 
the quantum system as the coupling is varied. Results are shown in Fig. \ref{figexpectation}.

\vspace{1.0 cm}

In general, agreement with the classical result is found: As the threshold coupling $\alpha=0.866$ (for $k=-1$) 
and $\alpha=0.5$ (for $k=0$) is crossed, the maximal possible number of molecules that can be formed for each 
case ($100\%$ for $k=-1$ and $50\%$ for $k=0$) is reached. In both cases, there is an abrupt change in the 
expectation value $3\langle N_c\rangle/N$ at the threshold point. However, for $k=0$, the expectation 
value $3\langle N_c\rangle/N$ does not exhibit any sudden change, indicative of the fact that there 
is no boundary in Fig. \ref{fig:2}(c). Therefore, qualitative changes are observed between the cases of $k$ 
negative and zero and the case of $k$ positive.

\subsection{Quantum phase transitions}
In order to gain a better insight into the effect of the threshold couplings for the quantum system, in our final 
analysis we investigate the existence of quantum phase transitions in our model (\ref{hamreduzido}).

A Quantum Phase Transition (QPT) is usually defined as a phase transition in the ground-state of the system under 
the variation of some parameter. Basically, there is a sudden change in the structure of the ground state at the QPT, 
and the properties such as entanglement, correlations, etc reflect this sudden change \cite{ssadchev}. 
There are different methods to determine a QPT. In particular, we will study the behaviour of the energy gap 
and fidelity of the system to identify a QPT. Here we mention that a QPT is rigorously defined in the 
thermodynamical limit $N \rightarrow \infty$. For large but finite $N$ the system does display an 
increasing sharp distinction between ground state regions, called Quantum Pre-Phase Transitions (QPPT). 
The occurrence of a QPPT in a finite system is a precursor for a QPT in the thermodynamic limit. 
Let us now study the QPPT of the Hamiltonian (\ref{hamreduzido}).

\subsubsection{Energy gap}
One possibility to identify a QPPT is through the energy gap, which is defined as the difference 
between the first excited state and the ground-state of the system.
$$\Delta E = E_1-E_0. $$

\begin{figure}[htb]
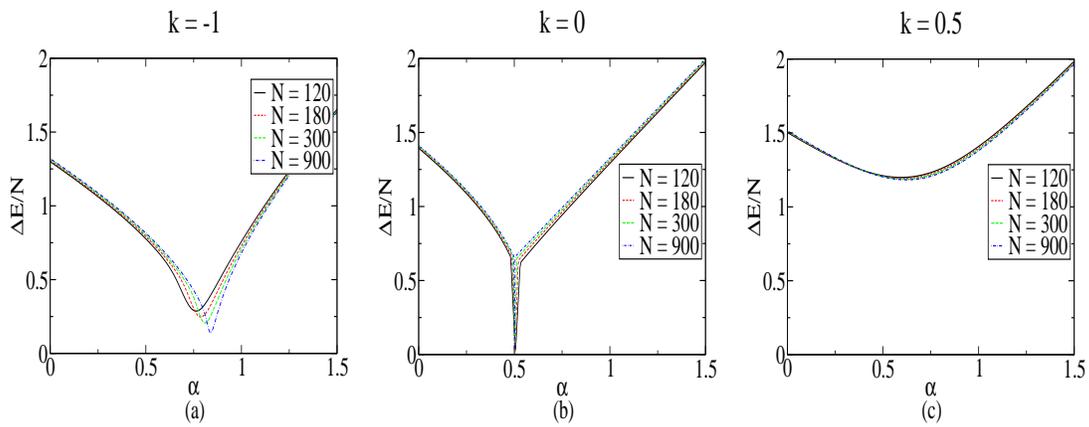

\begin{centering}
\epsfig{file=gapkneg1.eps,width=4.5cm,height=5.5cm,angle=0}\hspace{0.4cm}\epsfig{file=gapk0.eps,width=4.5cm,height=5.5cm,angle=0}\hspace{0.4cm}\epsfig{file=gapk05.eps,width=4.5cm,height=5.5cm,angle=0}
\par\end{centering}
\caption{(Color online) Energy gap between the first excited state and the ground state as a function of $\alpha$ for (a)$k=-1$; (b)$k=0$; (c)$k=0.5$ and different values of $N$. We are using $\Omega=1$ and natural units.}
\label{gap} 
\end{figure}

In Fig. \ref{gap} we plot the gap against the coupling parameter $\alpha$ for the cases of $k=-1, 0$ and $0.5$ 
using $\Omega=1$ and different values of $N$. In all cases the energy gap exhibits a minimum, 
which is much more pronounced in the cases of $k$ negative and zero compared to the case where $k$ is positive. We observe that as long as $N$ increases, the point where the gap tends to vanish corresponds to $\alpha=0.866$ for $k=-1$ (Fig. \ref{gap}(a)) and $\alpha=0.5$ for $k=0$ (Fig. \ref{gap}(b)), in agreement with the classical analysis. In contrast, when $k$ is positive there is no abrupt variation of the energy gap as shown in Fig. \ref{gap}(c) and QPPT are not expected. 

\vspace{1.0 cm}

\subsubsection{Fidelity}
\begin{figure}[htb]
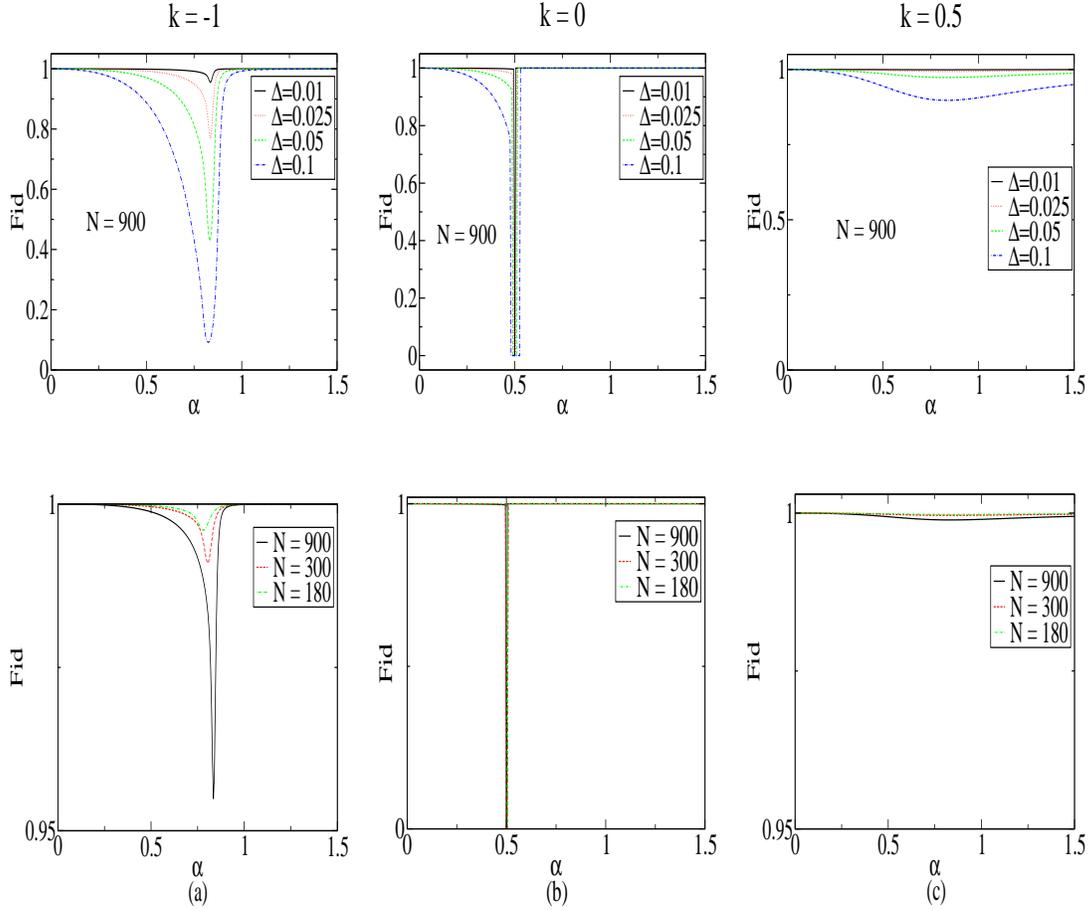

\begin{centering}
\epsfig{file=fidkneg1.eps,width=4.5cm,height=5.5cm,angle=0}\hspace{0.4cm}\epsfig{file=fidk0.eps,width=4.5cm,height=5.5cm,angle=0}\hspace{0.4cm}\epsfig{file=fidk05.eps,width=4.5cm,height=5.5cm,angle=0} \\ 
\vspace{1.0 cm}
\epsfig{file=fidkneg1N.eps,width=4.5cm,height=5.5cm,angle=0}\hspace{0.4cm}\epsfig{file=fidk0N.eps,width=4.5cm,height=5.5cm,angle=0}\hspace{0.4cm}\epsfig{file=fidk05N.eps,width=4.5cm,height=5.5cm,angle=0}
\par\end{centering}
\caption{(Color online) Ground-state wavefunction overlaps as a function of the coupling parameter $\alpha$ for (a)$k=-1$; (b)$k=0$; (c)$k=0.5$ and $\Omega=1$. On the top we are using $N=900$ and different values of $\Delta$. In the bottom we are using $\Delta=0.01$ and different values of $N$. In all cases the fidelity exhibits a minimum, which is substantially more pronounced for $k=-1$ and $k=0$, compared to $k=0.5$.}
\label{fid} 
\end{figure}

Another possibility to investigate the QPPT is through the behaviour of the fidelity, which is a concept widely used 
in the Quantum Information Theory \cite{nielsen,zhou2,buonsante}. The fidelity is basically defined as the modulus of 
the wavefunction overlap between two quantum states. Assuming the ground state of the system is non-degenerate, 
let $\Psi (\alpha)$
denote the unique normalized ground state. For fixed small $\Delta$ we define the function fidelity or 
ground-state wavefunction overlap $Fid_{\Delta}(\alpha)$ by
$$Fid_{\Delta}(\alpha) =|\langle \Psi(\alpha(1-\Delta))|\Psi(\alpha(1+\Delta))\rangle|,$$ 
which is symmetric in $\Delta$, bounded between $0$ and $1$ and satisfies 
$Fid_0(\alpha)=1$.
For systems which exhibit a quantum phase transition in the thermodynamic limit, the wavefunction overlaps 
between states in different phases go to zero in this limit. The occurrence of a minimum in the ground-state wavefunction 
overlap in a finite system is then a precursor for a quantum phase transition in the thermodynamic limit. 
Thus for finite systems we identify quantum phase \textit{pre-transitions} at a coupling $\alpha$ for which the 
fidelity is (locally) minimal. Fig \ref{fid} shows the behaviour of the fidelity 
for (a) $k=-1$; (b) $k=0$; (c) $k=0.5$ for fixed $N(\Delta)$ and different values of $\Delta(N)$ on 
the top (bottom). It is clear that the minimum value of $Fid_{\Delta}(\alpha)$, which determines 
the quantum phase pre-transition, tends to occur at $\alpha \approx0.86$ for $k=-1$ 
and $\alpha \approx 0.5$ for $k=0$. The distinction between the predicted threshold coupling 
and the observed pre-transition coupling is that the pre-transition coupling also occurs for $k$ 
positive, although for fixed $N$ the minimum of $Fid_{\Delta}(\alpha)$ is substantially more pronounced
for $k$ negative and zero compared to $k$ positive. 
In all instances the value of minimum decreases with increasing $N$. 
We remark that the value of $\alpha$ at which the minimum occurs 
is independent of $\Delta$, as shown in Fig.\ref{fid} (on the top). 
In our previous classical analysis qualitative differences are only found precisely when $k$ 
is negative or zero. Here it is clear that the distinguishability of two phases is more reliable also
for $k$ negative or zero.
We then interpret these results as the emergence of quantum phase boundaries for $k$ negative or zero.

\vspace{1.0 cm}

\section{Conclusion}
We have considered a model describing a mixture of two species of atoms in different proportions 
which can combine to form a bound molecular state at zero temperature. This hetero-triatomic molecular 
Bose-Einstein condensate model has been investigated in detail through a classical and a quantum analysis.

We have found that the fractional atomic population imbalance $k$, an extra control "knob" characteristic to heteronuclear 
models, plays an important role in the determination of the phase boundaries in the diagram of parameters in the classical 
analysis. This property also holds at a quantum level by inspecting the ground-state expectation values and the 
character of the quantum dynamics of the model.
We have also looked  for the quantum phase pre-transitions in our system and    
shown that the quantities energy gap and ground state fidelity are suited for revealing  
QPPT and pinning down the critical(bifurcation) points.

\section*{Acknowledgements}

C.C.N. K and A.F thank J. Links and E. Mattei for discussions. A. P. T., A.F., I. R. and Z. V. S. T. would like to thank CNPq - Conselho Nacional de Desenvolvimento Cient\'{\i}fico e Tecnol\'ogico for financial support, I.R. also thanks FAPERJ - Funda\c c\~ao Carlos Chagas Filho de Amparo \`a  Pesquisa do Estado do Rio de Janeiro for financial support. G. S. would like to thank CAPES - Coordena\c{c}\~ao de Aperfei\c{c}oamento de Pessoal de N\'{\i}vel Superior for financial support.


\end{document}